\documentclass[twocolumn,aps,prl,amsmath,amssymb]{revtex4-1}
\usepackage{graphicx,bm,amsmath,amssymb,natbib, verbatim,xcolor,gensymb}

\newcommand{\parallelsum}{\mathbin{\!/\mkern-5mu/\!}}

\begin{document}

\title{Spin-orbit splitting of Andreev states revealed by microwave spectroscopy}
\author{L. Tosi, C. Metzger, M. F. Goffman, C. Urbina, and H. Pothier}
\email[Corresponding author~: ]{hugues.pothier@cea.fr}
\affiliation{Quantronics group, Service de Physique de l'\'Etat Condens\'e (CNRS,
UMR\ 3680), IRAMIS, CEA-Saclay, Universit\'e Paris-Saclay, 91191 Gif-sur-Yvette, France}
\author{Sunghun Park}
\affiliation{Departamento de F\'{\i}sica Te\'orica de la Materia Condensada, Condensed Matter Physics Center (IFIMAC) and
Instituto Nicol\'as Cabrera, Universidad Aut\'onoma de Madrid, Spain}
\author{A. Levy Yeyati}
\affiliation{Departamento de F\'{\i}sica Te\'orica de la Materia Condensada, Condensed Matter Physics Center (IFIMAC) and
Instituto Nicol\'as Cabrera, Universidad Aut\'onoma de Madrid, Spain}
\author{J. Nyg\r{a}rd\textsuperscript{1} and P. Krogstrup\textsuperscript{1,2}}
\affiliation{\textsuperscript{1} Center for Quantum Devices,\\ \textsuperscript{2} Microsoft Quantum Materials Lab,\\ Niels Bohr Institute, University of Copenhagen, Copenhagen, Denmark}

\date{\today}

\begin{abstract}
We have performed microwave spectroscopy of Andreev states in superconducting weak links tailored in an InAs-Al (core-full shell) epitaxially-grown nanowire. The spectra present distinctive features, with bundles of four lines crossing when the superconducting phase difference across the weak link is 0 or $\pi.$ We interpret these as arising from zero-field spin-split Andreev states. A simple analytical model, which takes into account the Rashba spin-orbit interaction in a nanowire containing several {transverse subbands}, explains these features and their evolution with magnetic field. Our results show that the spin degree of freedom is addressable in Josephson junctions, and constitute a first step towards its manipulation.
\end{abstract}

\maketitle

\noindent\textbf{Introduction.} The Josephson supercurrent that flows through a weak link between two superconductors is a direct and generic manifestation of the coherence of the many-body superconducting state. The link can be a thin insulating barrier, a small piece of normal metal, a constriction or any other type of coherent conductor, but regardless of its specific nature the supercurrent is a periodic function of the phase difference $\delta$ between the electrodes \cite{Golubev2004}. However, the exact function is determined by the geometry and material properties of the weak link. A unifying microscopic description of the effect has been achieved in terms of the spectrum of discrete quasiparticle states that form at the weak link: the Andreev bound states (ABS) \cite{Kulik1970,Beenakker1991a,Furusaki1991,Bagwell1992}. The electrodynamics of an arbitrary Josephson weak link in a circuit is not only governed by the phase difference but depends also on the occupation of these states. Spectroscopy experiments on carbon nanotubes \cite{Pillet2010}, atomic contacts \cite{Bretheau2013,Bretheau2013_2,Janvier2015} and semiconducting nanowires \cite{Woerkom2016,Lee2014,Hays2017} have clearly revealed these fermionic states,
each of which can be occupied at most by two quasiparticles. The role of spin in these excitations is a topical issue in the rapidly growing fields of hybrid superconducting devices \cite{Michelsen2008,DeFranceschi2010,Linder2015} and of topological superconductivity \cite{Prada2017,Zazunov2017,Deng2018,Hart2017}. It has been predicted that for finite-length weak links the combination of a phase difference, which breaks time-reversal symmetry, and of spin-orbit coupling, which breaks spin-rotation symmetry, is enough to lift the spin degeneracy, giving therefore rise to spin-dependent Josephson supercurrents even in the absence of an external magnetic field \cite{Chtchelkatchev2003,Padurariu2010,Beri2008,Cayao2015}. Here we report the first observation of transitions between zero-field spin-split ABS.

\noindent\textbf{ABS and spin-orbit interaction}. Andreev bound states are formed from the coherent Andreev reflections that quasiparticles undergo at both ends of a weak link. Quasiparticles acquire a phase at each of these Andreev reflections and while propagating along the weak link of length $L.$ Therefore, the ABS energies depends on $\delta,$ on the transmission probabilities for electrons through the weak link and on the ratio $\lambda=L/\xi$ where $\xi$ is the superconducting coherence length. Assuming ballistic propagation, $\xi=\hbar v_F/\Delta$ is given in terms of the velocity $v_F$ of quasiparticles at the Fermi level within the weak link and of the energy gap $\Delta$ of the superconducting electrodes.
In a short junction, defined by $L \ll \xi,$ each conduction channel of the weak link, with transmission probability $\tau,$ gives rise to a single spin-degenerate Andreev level at energy $E_{A}=\Delta\sqrt{1-\tau\sin^{2}\left(\delta/2\right)}$ \cite{Beenakker1991a,Furusaki1991,Bagwell1992}. 
This simple limit has been probed in experiments on aluminum superconducting atomic contacts, using three different methods: Josephson spectroscopy \cite{Bretheau2013}, switching current spectroscopy \cite{Bretheau2013_2} and microwave spectroscopy in a circuit-QED setup \cite{Janvier2015}. The spectrum of Andreev states in a weak link with a sizable spin-orbit coupling has already been probed in two experiments on InAs nanowires \cite{Woerkom2016,Hays2017}. Both experiments were performed in the limit $L\lesssim \xi$. In Ref.~\cite{Hays2017}, the zero-field spectrum was probed using a circuit-QED setup and no effect of spin-orbit interaction was reported. In Ref.~\cite{Woerkom2016}, where spectra at finite field were obtained by Josephson spectroscopy, spin-orbit interaction enters in the interpretation of the spectra when the Zeeman energy is comparable to the superconducting gap \cite{Heck2017}.

\begin{figure*}[!t]
\includegraphics[width=2\columnwidth]{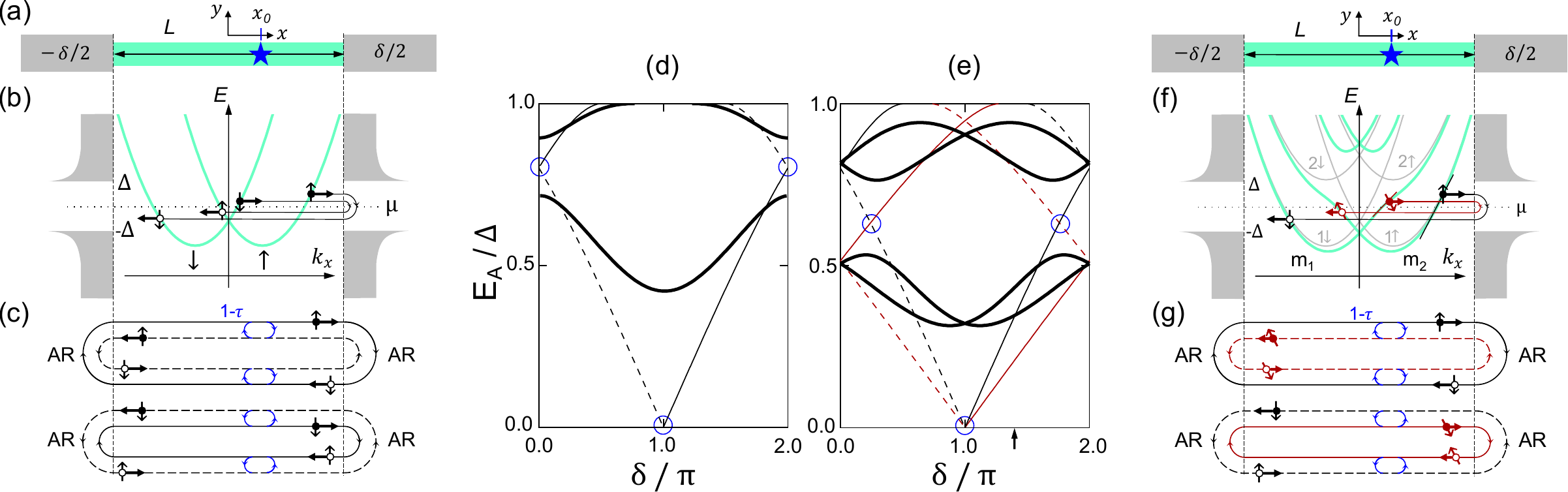}
\caption{{Effect of the Rashba spin-orbit coupling (RSO) on Andreev levels. 
(a) Weak link of length $L$ between superconductors with phase difference $\delta$. Blue star symbolizes a scatterer at position $x_0$. 
(b) Dispersion relation for a purely one-dimensional weak link in the presence of RSO (green solid lines, labels $\uparrow\downarrow$ indicate spin in y-direction). Density of states of superconducting electrodes is sketched at both ends of the wire. 
(c) Andreev reflections (AR) at the superconductors couples electrons (full circles) with holes (open circles) of opposite spins and velocities, leading to the formation of ABS. Blue arrows indicate reflections due to a scatterer.
(d) Energy of ABS (excitation representation). Thin lines in (c) and (d): transmission $\tau=1$, ABS formed from right-moving electrons and left-moving holes (solid) or the opposite (dashed). Backscattering ($\tau\ne1$) leads to opening of gaps at the crossings highlighted with blue circles in (d). Resulting spin-degenerate Andreev levels are shown with thick solid lines.
(e-g) Effect of RSO in the presence of two transverse subbands, only the lowest one being occupied. 
(f) Dispersion relation (subband spacing and superconducting gap are in a ratio that roughly corresponds to our experiments). Grey solid lines labelled 1$\uparrow\downarrow$ and 2$\uparrow\downarrow$ are dispersion relations for uncoupled subbands. RSO couples states of different subbands and opposite spins, leading to hybridized bands (green solid lines) with energy-dependent spin textures. Fermi level $\mu$ is such that only the lowest energy bands $m_1$ and $m_2$ are occupied. AR couples for example a fast electron from $m_2$ to a fast hole (in black), and a slow electron from $m_1$ to a slow hole (in red). 
(g) Construction of ABS: black and red loops are characterized by different absolute velocities. Spins pointing in different directions symbolize spin textures of the bands.
Thin red and black lines, solid and dashed in (e,g): ABS at $\tau=1$, associated with different spin textures. Thick black lines in (e): ABS when crossings highlighted with blue circles are avoided due to backscattering.}
}
\label{Fig:Rashba}
\end{figure*}

\begin{figure}[!t]
\includegraphics[width=0.8\columnwidth]{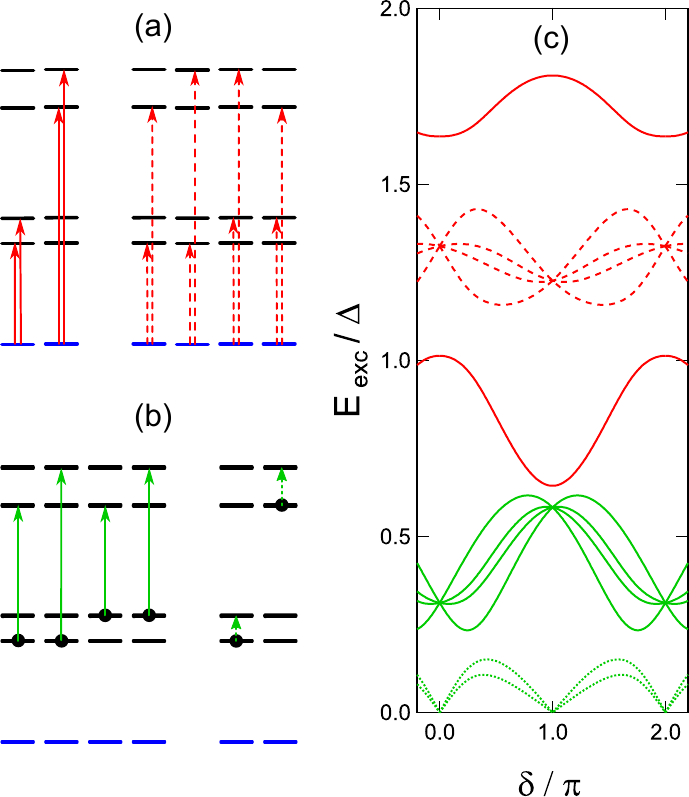}
\caption{{Possible parity-conserving transitions in a weak link with spin-split ABS (levels positions correspond to the phase indicated with an arrow in Fig.~\ref{Fig:Rashba}(e)). Blue line corresponds to the ground state. (a) {\it Pair} transitions. A pair of quasiparticles are created from the ground state, either both in the same manifold (solid arrows) or not (dashed arrows). (b) {\it Single-particle} transitions. A quasiparticle already present in one ABS (solid dot) is excited to another ABS, either in the same (dotted arrows)  or in another (solid arrows) manifold. (c) Corresponding transition energies, as a function of the phase difference $\delta$ across the weak link. (Transitions involving quasiparticles in the continuum are not represented).}
}
\label{Fig:Rashba2}
\end{figure}

{In the following, we consider a finite-length weak link with Rashba spin-orbit interaction (Fig.~\ref{Fig:Rashba}(a)), and show that spin-split Andreev states require at least two transverse subbands. We first discuss the case of a purely one-dimensional weak link. As shown by the green lines in Fig.~\ref{Fig:Rashba}(b) spin-orbit interaction splits the dispersion relation (assumed to be parabolic) according to the electron spin direction \cite{Bychkov1984}. Andreev reflections (AR) at the superconductors couples electrons (full circles) with holes (open circles) of opposite spins and velocities. When the transmission probability across the wire is perfect $(\tau=1)$, Andreev bound states arise when the total accumulated phase along closed paths that involve two AR and the propagation of an electron and a hole in opposite directions (Fig.~\ref{Fig:Rashba}(c)) is a multiple of $2\pi$ \cite{Kulik1970}. Figure 1(d) shows, in the excitation representation, the energy of the resulting ABS as a function of $\delta$. ABS built with right- (left-) moving electrons are shown with thin solid (dashed) lines in Fig.~\ref{Fig:Rashba}(c\&d). Note that the existence of two ABS at some phases is just a finite length effect \cite{Bagwell1992} (here $L/\xi=0.8$) and that ABS remain spin-degenerate as the spatial phase acquired by the electron and the Andreev-reflected hole is the same for both spin directions. Backscattering in the weak link ($\tau\ne1$), due either to impurities or to the spatial variation of the electrostatic potential along the wire, couples electrons (as well as holes) of the same spin travelling in opposite directions, leading to avoided crossings at the points indicated by the open blue circles in Fig. \ref{Fig:Rashba}(d). One obtains in this case two distinct Andreev states (thick solid lines), which remain spin-degenerate. 
This is no longer the case in the presence of a second transverse subband, even if just the lowest one is actually occupied \cite{Reynoso2012, Yokoyama2014, Murani2016, Park2017}. Figure~\ref{Fig:Rashba}(f) shows how spin-orbit coupling hybridizes the spin-split dispersion relations of the two subbands (around the crossing points of 1$\uparrow$ with 2$\downarrow$ and of 1$\downarrow$ with 2$\uparrow$)\cite{Moroz1999,Governale2002}. The new dispersion relations become non-parabolic and are characterized by different energy-dependent spin textures \cite{Moroz1999,Governale2002,Reynoso2012,Yokoyama2014, Murani2016, Park2017}. We focus on a situation in which only the two lowest ones ($m_1$ and $m_2$ in the figure) are occupied. Importantly, their associated Fermi velocities are different. When $\tau=1$, this leads, as illustrated by Fig.~\ref{Fig:Rashba}(e,g), to two families of ABS represented by black and red thin lines, built from states with different spin textures. As before, backscattering leads to avoided crossings at the points indicated by the blue open circles in Fig.~\ref{Fig:Rashba}(e). The resulting ABS group in manifolds of spin-split states represented by the thick black lines. In the absence of a magnetic field, the states remain degenerate at $\delta = 0$ and $\pi$.} 
{Figure~\ref{Fig:Rashba2} shows parity-conserving transitions that can be induced by absorption of a microwave photon, at a given phase. Red arrows (Fig.~\ref{Fig:Rashba2}(a)) correspond to {\it pair} transitions in which the system is initially in the ground state and a pair of quasiparticles is created, either in one manifold or in different ones. Green arrows (Fig.~\ref{Fig:Rashba2}(b)) correspond to {\it single-particle} transitions where a trapped quasiparticle \cite{Zgirski2011} already occupying an Andreev state is excited to another one \cite{Vayrynen2015,Park2017} which can be in the same or in another ABS manifold. The corresponding transition energies in the absorption spectrum for both the pair and single-particle cases are shown in Fig.~\ref{Fig:Rashba2}(c), as a function of the phase difference $\delta$. Pair transitions that create two quasiparticles in the same energy manifold do not carry information on the spin structure. On the contrary, pair and single-particle transitions  involving different energy manifolds produce peculiar bundles of four distinct lines all crossing at $\delta=0$ and $\delta=\pi$. They are a direct signature of the spin splitting of ABS. Finally, single-particle transitions within a manifold give rise to bundles of 2 lines. As discussed below, some of these transitions are accessible in our experiment.}
\begin{figure}[]
\includegraphics[width=0.8\columnwidth]{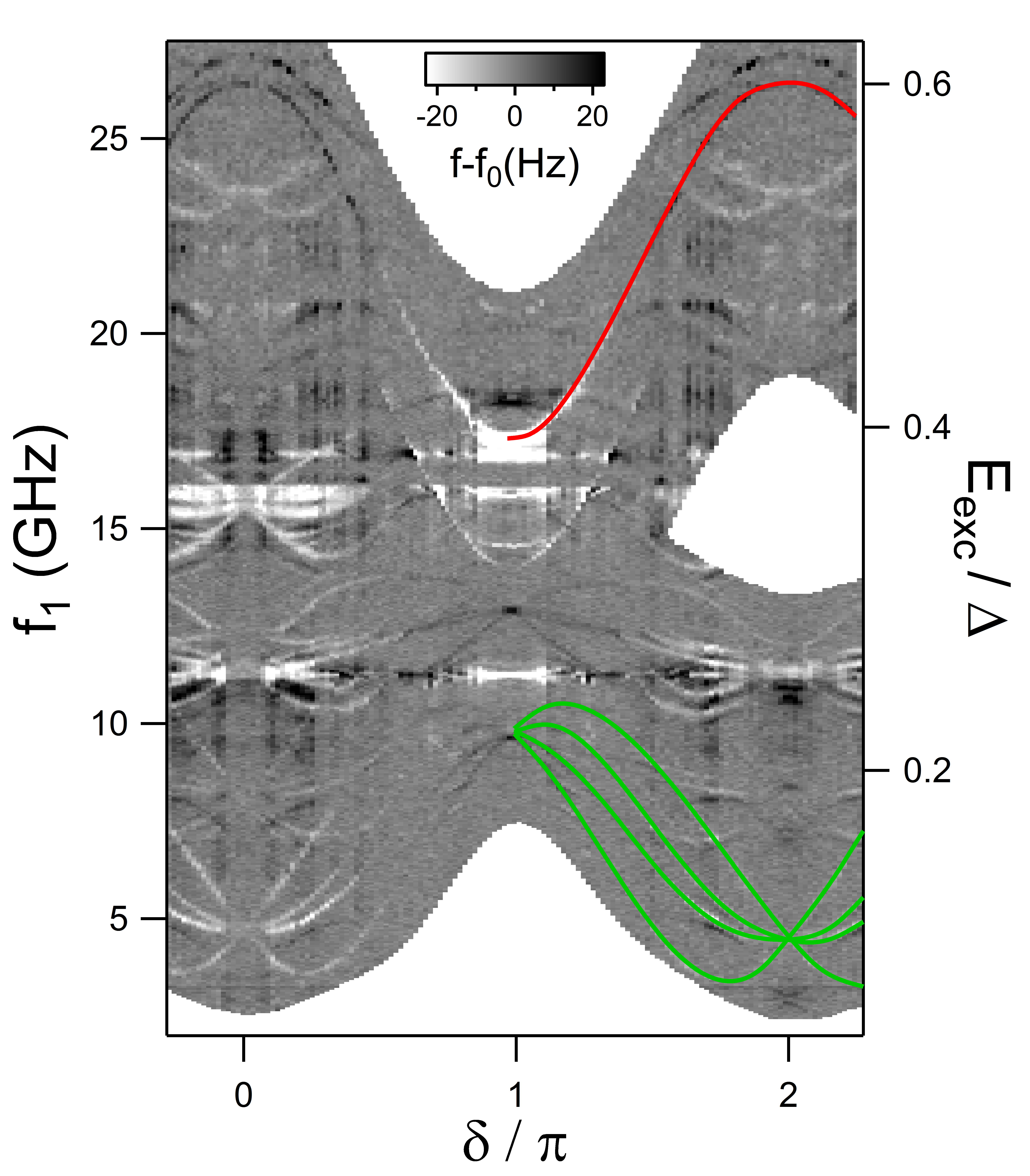}
\caption{Microwave excitation spectrum measured at a gate voltage $V_g=-0.89~$V. The grey scale represents the frequency change $f-f_0$ of a resonator coupled to the weak link when a microwave excitation at frequency $f_1$ is applied, as a function of the phase difference $\delta$ across the weak link. In the right half of the figure, some transition lines are highlighted. {Red line corresponds to a pair transition, green lines are single-particle transitions}.}\label{Fig2:SpectrumZeroField}
\end{figure}

Figure \ref{Fig2:SpectrumZeroField} shows a spectrum measured on an InAs nanowire weak link between aluminum electrodes. The plot shows at which frequency $f_1$ microwave photons are absorbed, as a function of the phase difference $\delta$ across the weak link (see description of the experiment below). This is a very rich spectrum, but here we point two salient features highlighted with color lines on the right half side of the figure. The red line corresponds to {a pair transition}, with extrema at $\delta=0$ and $\delta=\pi.$ The frequency $f_1(\delta=0)=26.5~\rm{GHz}$ is much smaller than twice the gap of aluminum $2\Delta/h=88~$GHz, as expected for a junction longer than the coherence length. To the best of our knowledge, this is the first observation of a discrete Andreev spectrum in the long-junction limit. The observation of the bundle of lines (in green) with crossings at $\delta=0$ and $\delta=\pi$ that clearly corresponds to {single-particle transitions} shown in Fig. \ref{Fig:Rashba2}(c) is the central result of this work. 
\begin{figure}[ht]
\includegraphics[width=\columnwidth]{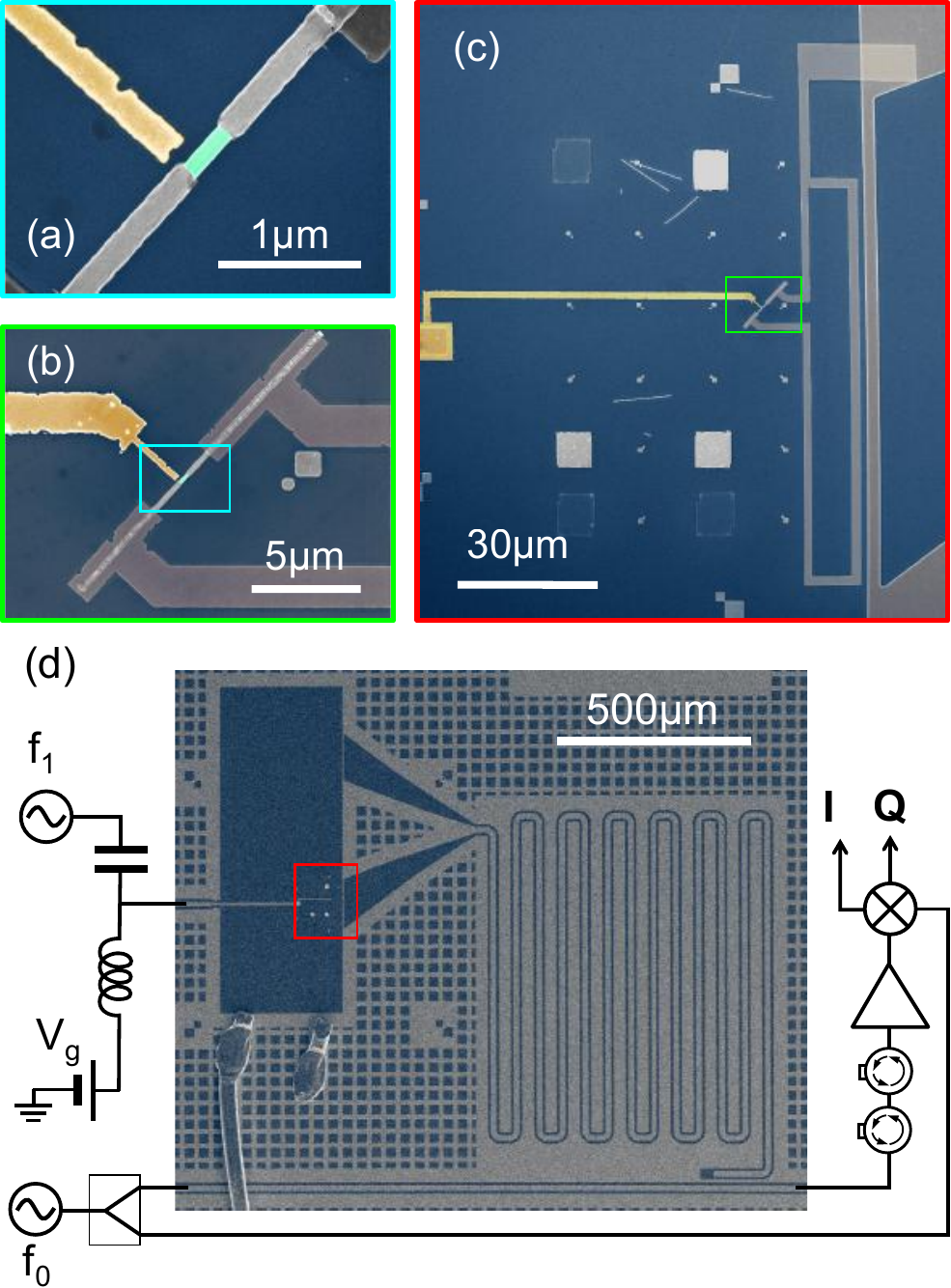}
\caption{Experimental setup. (a) False-color scanning electron microscope image of the InAs-Al core-shell nanowire. The Al shell (grey) was removed over 370~nm to form the weak link between the superconducting electrodes. A close-by side electrode (Au, yellow) is used to gate the InAs exposed region (green). (b),(c) The nanowire is connected to Al leads that form a loop. This loop is located close to the shorted end of a coplanar wave guide (CPW) resonator. (d) The CPW resonator is probed by sending through a bus line a continuous microwave tone at its resonant frequency $f_0=3.26$~GHz and demodulating the transmitted signal, yielding quadratures $I$ and $Q$. Microwaves inducing Andreev transitions are applied through the side gate (frequency $f_1$) using a bias tee, the DC port being used to apply a DC voltage $V_g$.}
\label{Fig1:Setup}
\end{figure}

\begin{figure*}[t!]
\includegraphics[width=1.6\columnwidth]{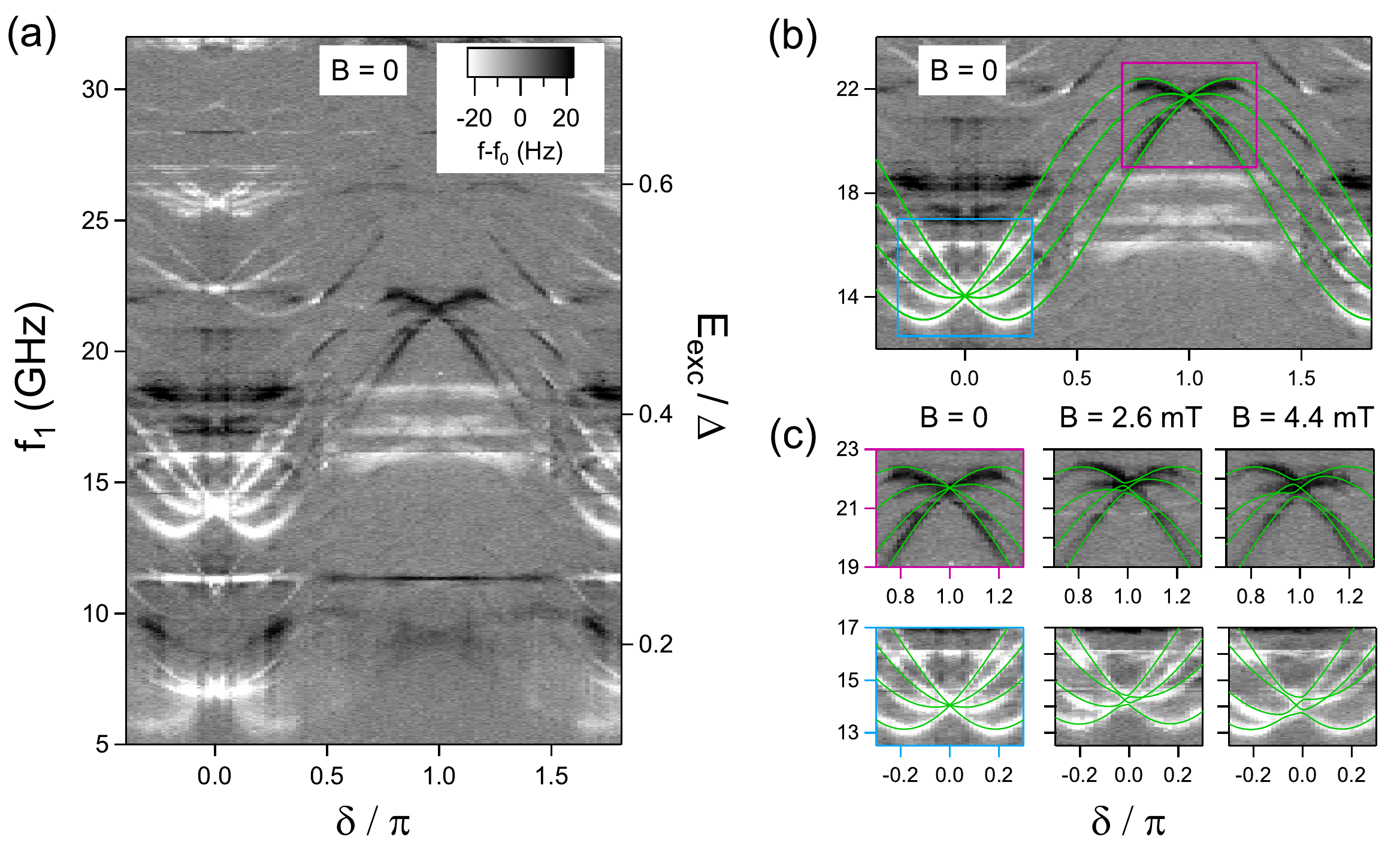}
\caption{Excitation spectra at $V_g=0.5$~V. (a) Large scale spectrum at zero magnetic field. (b) Zoom on the same data, with fits (see text). (c) Dependence of the spectrum with the amplitude $B$ of an in-plane magnetic field applied at an angle of $-45\degree$ with respect to the nanowire axis. Green lines are fits (see text).}
\label{Fig:Spectrum2}
\end{figure*}

\noindent \textbf{Experimental setup}. The measurements are obtained using the circuit-QED setup shown in Fig.~\ref{Fig1:Setup}(d), and performed at $\sim 40$~mK in a pulse-tube dilution refrigerator. The superconducting weak link was obtained by etching away, over a 370-nm-long section, the 25-nm-thick aluminum shell that fully covers a 140-nm-diameter InAs nanowire \cite{Krogstrup2015,Chang2015,Goffman2017} (see Figs.~\ref{Fig1:Setup}(a) and (b)). A side-gate allows tuning the charge carrier density and the electrostatic potential in the nanowire and therefore the Andreev spectra \cite{Woerkom2016}. The weak link is part of an aluminum loop of area $S\sim 10^3~\mu \rm{m}^2$, which has a connection to ground to define a reference for the gate voltage (see Fig.~\ref{Fig1:Setup}(c)). The phase $\delta$ across the weak link is imposed by a {small} magnetic field $B_z~(<5\mu \rm{T})$ perpendicular to the sample plane: $\delta=B_z S/\varphi_0$, with $\varphi_0=\hbar/2e$ the reduced flux quantum. Two additional coils are used to apply a magnetic field in the plane of the sample.  The loop is inductively coupled to the shorted end of a $\lambda/4$ microwave resonator made out of Nb, with resonance frequency $f_0\approx3.26~\rm{GHz}$ and internal quality factor $Q_{\rm{int}}\approx 3\times10^5.$ A continuous signal at frequency $f_0$ is sent through a coplanar transmission line coupled to the resonator (coupling quality factor $Q_{\rm{c}}\approx 1.7\times10^5$), and the two quadratures $I$ and $Q$ of the transmitted signal are measured using homodyne detection (see Fig.~\ref{Fig1:Setup}(d)). Andreev excitations in the weak link are induced by a microwave signal of frequency $f_1$ applied on the side gate. The corresponding microwave source is chopped at 3.3~kHz, and the response in $I$ and $Q$ is detected using two lock-ins, with an integration time of $0.1~\rm{s}.$ {This response is expressed in terms of the corresponding frequency shift $f-f_0$ in the resonator (see Appendix A3).} The fact that {single-particle} transitions are observed (see Fig.~\ref{Fig2:SpectrumZeroField}) means that during part of the measurement time Andreev states are occupied by a single quasiparticle. This is in agreement with previous experiments, in which the fluctuation rates for the occupation of Andreev states by out-of-equilibrium quasiparticles were found to be in the 10~ms\textsuperscript{-1} range \cite{Zgirski2011,Janvier2015,Hays2017}. Note that in contrast to an excitation that couples to the phase across the contact {through the resonator} \cite{Janvier2015,Heck2017,Park2017}, exciting through the gate allows {to drive } transitions away from $\delta=\pi$ and at frequencies very far detuned from that of the resonator.

\noindent \textbf{Spectroscopy at zero magnetic field}. Figure~\ref{Fig:Spectrum2}(a) presents another spectrum taken at zero magnetic field (apart from the tiny perpendicular field $B_z<5~\mu\rm{T}$ required for the phase biasing of the weak link), at $V_g=0.5$~V. In comparison with the spectrum in Fig.~\ref{Fig2:SpectrumZeroField}, {pair transitions} are hardly visible in Fig.~\ref{Fig:Spectrum2}. Bundles of lines corresponding to  {single-particle transitions} have crossings at  7.1, 14.0 and 22.4~GHz at $\delta=0$ and 9, 21.5 and 26.0~GHz at $\delta=\pi$. Here, as in Fig.~\ref{Fig2:SpectrumZeroField} {(see Appendix A2)}, replicas of transition lines shifted by $f_0$ are also visible {(bundle of lines near $f_1=11$~GHz and around $\delta=0$)}. They correspond to transitions involving the absorption of a photon from the resonator. {Remarkably, the sign of the response appears correlated with the curvature of the transition lines. This suggests that the signal is mainly associated with a change in the effective inductance of the nanowire weak link. Additional work is needed to confirm this interpretation}. We focus on the bundle of lines between 13 and 23~GHz for which the effect of a magnetic field $B$ was also explored. 
Green lines in Fig.~\ref{Fig:Spectrum2}(b) are fits of the data at $B=0$ with a simple model that accounts for two bands with different Fermi velocities $v_1$ and $v_2$, and the presence of a single scatterer in the wire (see Appendix A1). The model parameters are $\lambda_{j=1,2}=L\Delta/(\hbar v_j)$ and the position $x_0 \in [-L/2,L/2]$ of the scatterer of transmission $\tau$. ABS are found at energies $E=\epsilon \Delta$, with $\epsilon$ solution of the transcendental equation (see Appendix A1):
\begin{eqnarray}
\tau\cos\left((\lambda_{1}-\lambda_{2})\epsilon \mp\delta\right) + (1-\tau) \cos((\lambda_{1}+\lambda_{2}) \epsilon x_r) &=& \nonumber \\
 \cos(2 \arccos \epsilon - (\lambda_{1}+\lambda_{2})\epsilon) 
 && \;
\label{single-barrier}
\end{eqnarray} 
where $x_r = 2x_0/L$. {It should be noticed that Eq.~(\ref{single-barrier}) for $\lambda_1=\lambda_2$ reduces to the known result for a single quantum channel without spin-orbit \cite{Bagwell1992,Samuelsson2000}.}  
The fit in Fig.\,\ref{Fig:Spectrum2}(b) corresponds to $\lambda_1=1.3,$ $\lambda_2=2.3,$ $\tau=0.295$ and $x_r=0.525$ (we take $\Delta=182\,\mu$eV$=h\times 44\,$GHz for the gap of Al). These values can be related to microscopic parameters, in particular to the intensity $\alpha$ of the Rashba spin-orbit interaction entering in the Hamiltonian of the system as $H_{R}=-\alpha (k_x \sigma_y-k_y \sigma_x)$ (with $\sigma_{x,y}$ Pauli matrices acting in the spin) \cite{Park2017}. Assuming a parabolic transverse confinement potential, an effective wire diameter of $W=140$\,nm, and an effective junction length of $L=370$\,nm, the values of $\lambda_{1,2}$ are obtained for $\mu=422\,\mu$eV (measured from the bottom of the band) and $\alpha=38$\,meV\,nm, a value consistent with previous estimations \cite{Fasth2007,Scherubl2016}. However, we stress that this estimation is model-dependent: very similar fits of the data can be obtained using a double-barrier model (with scattering barriers located at the left $(x=-L/2)$ and right $(x=L/2)$ edges of the wire) with $\lambda_1=1.1$ and  $\lambda_2=1.9,$ leading to $\alpha = 32$\,meV\,nm. For both models, we get only two manifolds of Andreev levels in the spectrum, and only these four {single-particle transitions} are expected in this frequency window (transitions within a manifold are all below 3.5\,GHz). The other observed bundles of transitions are attributed to other conduction channels: although we considered till now only one occupied transverse subband, the same effect of spin-dependent velocities is found if several subbands cross the Fermi level. A more elaborate model, together with a realistic modeling of the bands of the nanowire, is required to treat this situation and obtain a quantitative fit of the whole spectra.

\noindent \textbf{Spin character of ABS.} {The splitting of the ABS and the associated transitions in the absence of a Zeeman field reveal the difference in the Fermi velocities $v_1$ and $v_2$, arising from the spin-orbit coupling in the multi-channel wire. To further confirm that this is indeed a spin effect we probe} the ABS spectra under a finite magnetic field, and in particular as a function of the orientation of the field with respect to the nanowire axis \cite{Park2017}. Figure\,\ref{Fig:Spectrum2}(c) shows the spectrum in presence of an in-plane magnetic field with amplitudes $B=0,$ 2.6 and 4.4\,mT applied at an angle of $-45\degree$  with respect to the wire axis. The symmetry around $\delta=0$ and $\delta=\pi$ is lost. This is accounted for by an extension of the single-barrier model at finite magnetic field (green lines) {and assuming an anisotropic g-factor: $g_{\perp}=12$ and $g_{\parallelsum}=8$ (see below and Appendix A1).}

\begin{figure}[t!]
\includegraphics[width=0.8\columnwidth]{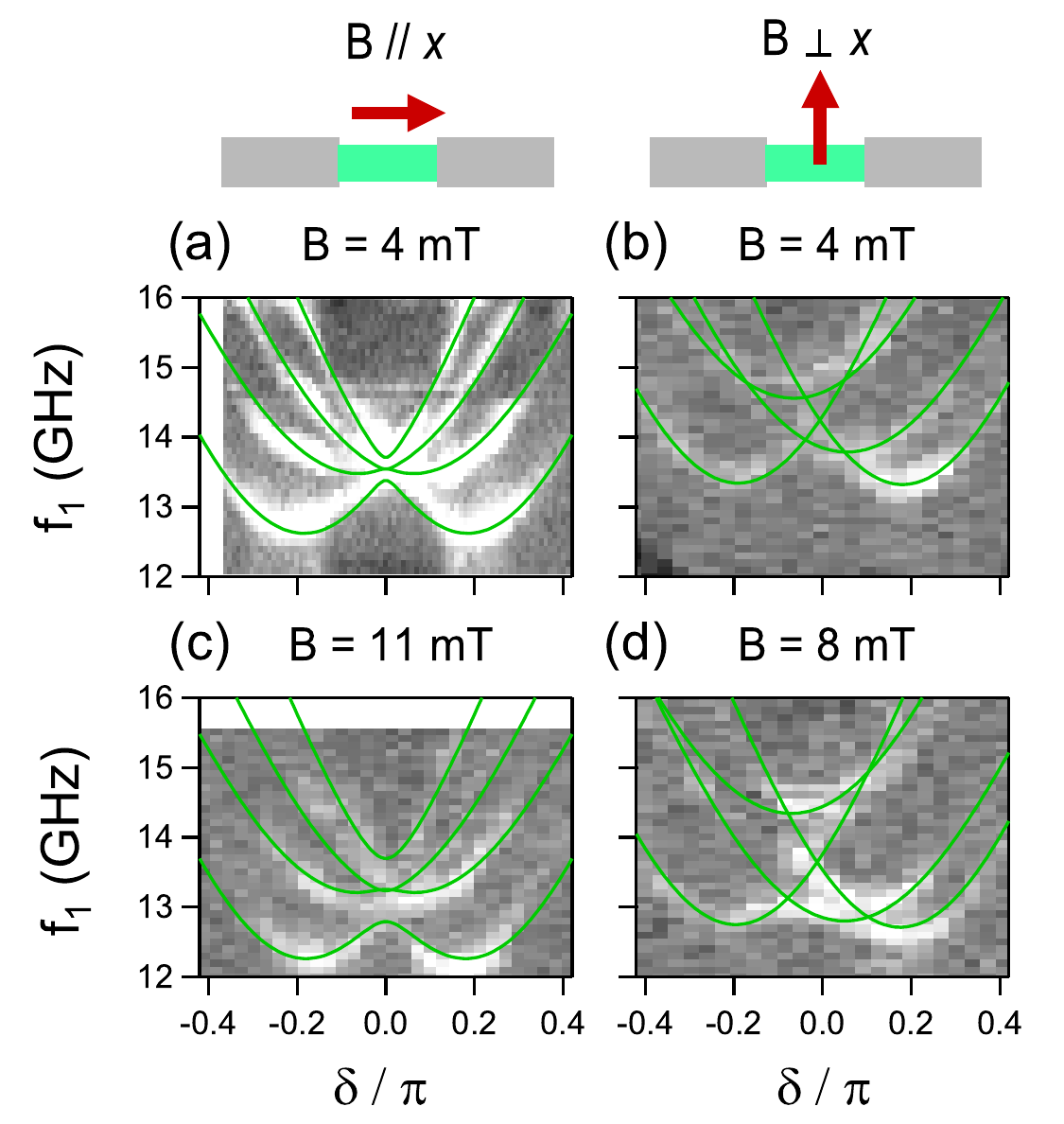}
\caption{Effect of an in-plane magnetic field on the ABS excitation spectrum around $\delta=0$. The Andreev states correspond to the same gate voltage as in Fig.\,\ref{Fig:Spectrum2}. Field is applied parallel (a,c) or perpendicular (b,d) to the wire. Green lines are the result from the theory, {using $g_{\perp}=12$ and $g_{\parallelsum}=8$ }(see text).}
\label{Fig:Field}
\end{figure}

The specific effects of a parallel and of a perpendicular magnetic field on the ABS are shown in Fig.\,\ref{Fig:Field}. When the field is perpendicular to the wire ($B\perp x$), the ABS spectrum becomes asymmetric (this is related to the physics of $\varphi_0$ junctions \cite{Reynoso2012}), as observed in Fig.\,\ref{Fig:Field}(b,d). The field is directly acting in the quantization direction of the spin-split transverse subbands (gray parabolas in Fig.\ \ref{Fig:Rashba}(f)) from which the ABS are constructed, leading to Zeeman shifts of the energies. When the field is along the wire axis $B \parallelsum x$, and thus perpendicular to the spin quantization direction, it mixes the spin textures and lifts partly the degeneracies at $\delta=0$ and $\delta=\pi$ (see Fig.\,\ref{Fig:FieldBands} in Appendix). The spectrum of ABS is then modified, but remains symmetric \cite{angle} around $\delta=0$ and $\pi,$ see Fig.\,\ref{Fig:Field}(a,c). 
{Keeping the same parameters as in Fig.\,\ref{Fig:Spectrum2}, the value of the g-factor was taken as a fit parameter for all the data with perpendicular field, and for all the data with parallel field, leading to two distinct values of the g-factor: $g_{\perp}=12$ and $g_{\parallelsum}=8$ (see Appendix). Green lines show the resulting best fits.}

\noindent\textbf{Concluding remarks}. The results reported here show that the quasiparticle spin can be a relevant degree of freedom in Josephson weak links, even in the absence of a magnetic field. 
This work leaves several open questions. 
{Would a more realistic modeling of the nanowire \cite{Degtyarev2017,Zuo2017,Antipov2018,Winkler2018} allow for a precise determination of spin-orbit interaction from the measured spectra?}
We need to understand, {along the lines of \cite{Gorelik1995} for example,} the coupling between the microwave photons and the ABS when the excitation is induced through an electric field modulation, as done here, instead of a phase modulation \cite{Desposito2001,Vayrynen2015,Park2017}. 
In particular, what are the selection rules? Are transitions between ABS belonging to the same manifold allowed? Can one observe {pair transitions} leading to states with quasiparticles in different manifolds? What determines the signal amplitude?
Independently of the answer to these questions, the observation of spin-resolved transitions between ABS constitutes a first step towards the manipulation of the spin of a single superconducting quasiparticle \cite{Chtchelkatchev2003,Park2017}. {Would} the spin coherence time of a localized quasiparticle be  different from that of a propagating one \cite{Quay2015}? Finally, we think that the experimental strategy used here could allow probing a topological phase with Majorana bound states at larger magnetic fields \cite{Vayrynen2015}.

\acknowledgments  
Technical support from P. S\'enat is gratefully acknowledged. {We thank M. Devoret, M. Hays and K. Serniak for sharing their results on a similar experiment and for discussions.}
We thank A. Reynoso {for providing  us codes related to his work \cite{Reynoso2012} and for useful discussions. We also acknowledge discussions with \c{C}. Girit, H. Bouchiat, A. Murani and our colleagues from the Quantronics group}. We thank P. Orfila and S. Delprat for nanofabrication support.
This work has been supported ANR contract JETS, by the Renatech network, by the Spanish MINECO through Grant No.~FIS2014-55486-P, ~FIS2017-84860-R 
and through the ``Mar\'{\i}a de Maeztu'' Programme for Units of Excellence in R\&D (MDM-2014-0377). Center for Quantum Devices is supported by the Danish National Research Foundation. 
L. Tosi was supported by the Marie Sk\l{}odowska-Curie individual fellowship grant 705467.
P. Krogstrup acknowledges support from Microsoft Quantum and the European Research Council (ERC) under the grant agreement No.716655 (HEMs-DAM)

\begin{widetext}
\appendix

\section{Appendix}

\subsection{A1. Details on the theoretical model and the fitting parameters}
\label{Sec:Appendix1}

{The nanowire is described by the Hamiltonian $H^{\text{3D}}$ consisting of a kinetic energy, a confining harmonic potential in $y$ and $z$-directions with a confinement width $W$ (effective diameter of the nanowire) and Rashba spin-orbit coupling with intensity $\alpha$,  
\begin{equation}
H^{\text{3D}} = \frac{p^{2}_x+p^{2}_y+p^{2}_z}{2 m^*} + \frac{ \hbar^2 (y^2 + z^2)}{2 m^* (W/2)^4} + \alpha (-k_x \sigma_y + k_y \sigma_x), 
\label{ModelHamiltonian}
\end{equation} 
where $m^*$ is the effective mass and $\sigma_{x,y}$ are the Pauli matrices for spin. We consider two spin-full} transverse subbands denoted by $n\sigma$, with $n=1,2$ and $\sigma=\uparrow,\downarrow$, arising from the confining potential in the transverse direction (gray parabolas in Fig.\,\ref{Fig:Rashba}({f})) under the effect of the Rashba spin-orbit coupling. The energy dispersion relations of the resulting lowest subbands (green lines labelled $m_1$ and $m_2$ in Fig. \ref{Fig:Rashba}({f})) are \cite{Park2017} 

\begin{equation}
E_s(k_x) = \frac{\hbar^2 k^2_x}{2 m^*} + \frac{E^{\perp}_1 +E^{\perp}_2}{2} 
- \sqrt{\left(  \frac{E^{\perp}_1 - E^{\perp}_2}{2} -s \alpha k_x\right)^2 + \eta^2}, 
\label{Dispersion}
\end{equation} 
where $s=-1$ corresponds to $m_1$ and $s=+1$ to $m_2$, and $E^{\perp}_n = 4\hbar^2 n/(m^* W^2)$. $\eta=\sqrt{2}\alpha/W$ is the strength of the subband mixing due to the Rashba spin-orbit coupling. {In accordance to the estimated nanowire diameter we take $W\sim 140$ nm, which leads to $E^{\perp}_2-E^{\perp}_1 \sim 0.68$ meV for the subband separation.}
Particle backscattering within the nanowire is accounted for by either a single delta-like potential barrier located at some arbitrary position $x_0$ or by potential barriers localized at both ends ($x=\pm L/2$).  

The linearized Bogoliubov-de Gennes equation around the chemical potential $\mu$ is
\begin{equation}
\begin{pmatrix}
H_0 + H_b & \Delta(x) e^{i \delta(x)}\\
\Delta(x) e^{-i \delta(x)} & -H_0 - H_b
\end{pmatrix} \Psi(x) = E_A \Psi(x)
\label{BdG}
\end{equation}
with the basis $\Psi(x)=(\psi^{e}_{+,R}(x),\psi^{e}_{+,L}(x),\psi^{e}_{-,R}(x),\psi^{e}_{-,L}(x),\psi^{h}_{+,R}(x),\psi^{h}_{+,L}(x),\psi^{h}_{-,R}(x),\psi^{h}_{-,L}(x))$, where $R(L)$ refers to the right-moving (left-moving) electron $(e)$ or hole $(h)$ in the bands $m_1$($-$), $m_2$($+$). Here $H_0$ is the Hamiltonian for electrons in the nanowire 
\begin{equation}
H_0=
\begin{pmatrix}
-i \hbar v_1 \partial_x - \hbar v_1 k_{F1} & 0 & 0 & 0 \\
 0 & i \hbar v_2 \partial_x - \hbar v_2 k_{F2} & 0 & 0 \\
 0 & 0 & -i \hbar v_2 \partial_x - \hbar v_2 k_{F2} & 0 \\ 
 0 & 0 & 0 & i \hbar v_1 \partial_x - \hbar v_1 k_{F1}
\end{pmatrix},
\end{equation}
where $v_{j=1,2}$ are the Fermi velocities given by 
\begin{equation}
v_j = \frac{\hbar k_{Fj}}{m^*} + (-1)^j \frac{\alpha\left(E^{\perp}_1/2-(-1)^j \alpha k_{Fj}\right)}{{\hbar} \sqrt{\left(E^{\perp}_1/2 - (-1)^j\alpha k_{Fj}\right)^2 + \eta^2}},
\label{Velocity}
\end{equation}
and $k_{Fj}$ are the Fermi wave vectors satisfying $E_s(k_{Fj})=\mu$. {We note that if there is no subband mixing, i.e. $\eta=0$ (gray parabolas in Fig.\,\ref{Fig:Rashba}(f)), Eqs. \eqref{Dispersion} and \eqref{Velocity} show that $k_{F1}-k_{F2}=2 m^* \alpha/ \hbar^2$ and $v_1 - v_2 = (k_{F1}-k_{F2}) \hbar/m^* - 2 \alpha/\hbar=0$, indicating clearly that the Fermi velocities are the same.}
The potential scattering term $H_b$ is modeled as 
\begin{equation}
H_b = U_b (x)  
 \begin{pmatrix}
1 & \text{cos}[(\theta_1-\theta_2)/2] & 0 & 0 \\
 \text{cos}[(\theta_1-\theta_2)/2] & 1 & 0 & 0 \\
 0 & 0 & 1 & \text{cos}[(\theta_1-\theta_2)/2] \\ 
 0 & 0 & \text{cos}[(\theta_1-\theta_2)/2] & 1
\end{pmatrix}, 
\end{equation}
where 
\begin{equation}
U_b(x) = 
\begin{cases}
U_0 \delta(x-x_0) & \text{for a single barrier at $x=x_0$} \\ 
U_L \delta(x+L/2) + U_R \delta(x-L/2) & \text{for barriers at $x=-L/2$ and $x=L/2$},
\end{cases}
\end{equation}
and $\theta_{j=1,2}=\arccos[(-1)^j (\hbar k_{F_j}/m^*-v_j)/\alpha]$ 
{characterize the mixing with the higher subbands, \textit{i.e.}  $\cos(\theta_j/2)$ and $\sin (\theta_j/2)$ determine the weight of the states on the hybridized subbands and therefore their spin texture.}
The superconducting order parameter $\Delta(x) e^{i \delta(x)}$ in Eq. \eqref{BdG} is given by $\Delta e^{-i \delta/2}$ at $x<-L/2,$ $\Delta e^{i \delta/2}$ at $x>L/2,$ and zero otherwise, where $\delta$ is the superconducting phase difference.

\textbf{Ballistic regime.} 
In the absence of particle backscattering, the phase accumulated in the Andreev reflection processes at $x=-L/2$ and $x=L/2$, as illustrated in Fig.~\ref{Fig:Rashba}(c), leads to the following transcendental equation for the energy $\epsilon = E_A/\Delta$ of the ABS as a function of $\delta:$ 
\begin{equation}
\sin(\epsilon\lambda_1 - s \delta/2 - \arccos \epsilon) \sin(\epsilon\lambda_2 +s \delta/2 - \arccos \epsilon) = 0,
\label{ballistic}
\end{equation}
where $\lambda_{j=1,2}=L\Delta/(\hbar v_j)$.
For $\epsilon \ll 1$, there are two sets of solutions given by 
\begin{eqnarray*}
\left\{ \begin{array}{l}
\epsilon_{\color{blue}{\uparrow}}(\delta) = \frac{1}{1+\lambda_1}\left[\frac{\delta}{2} + \left(l+\frac{1}{2}\right) \pi\right] \nonumber\\
\epsilon_{\color{blue}{\downarrow}}(\delta) = \frac{1}{1+\lambda_1}\left[-\frac{\delta}{2} + \left(l'+\frac{1}{2}\right) \pi\right] \nonumber \end{array}\right., \\
\left\{ \begin{array}{l}
\epsilon_{\color{blue}{\swarrow}}(\delta) = \frac{1}{1+\lambda_2}\left[\frac{\delta}{2} + \left(l+\frac{1}{2}\right) \pi\right] \nonumber\\
\epsilon_{\color{blue}{\nearrow}} (\delta)= \frac{1}{1+\lambda_2}\left[-\frac{\delta}{2} + \left(l'+\frac{1}{2}\right) \pi\right] \nonumber \end{array}\right.,\nonumber\\
\end{eqnarray*}
with integers $l$ and $l'$. The ballistic ABS are represented by the thin lines (black and red) in Fig.~\ref{Fig:Rashba}(e).

\textbf{Single barrier model.} 
In this case, the effect of the barrier can be taken into account as an additional boundary condition at $x=x_0$, 
\begin{equation}
\Psi(x_0+0^{+}) = 
\begin{pmatrix}
M_{12} & 0 & 0 & 0 \\
0 & M_{21} & 0 & 0 \\
0 & 0 & M_{12} & 0 \\
0 & 0 & 0 & M_{21}
\end{pmatrix}
\Psi(x_0-0^{+}),
\label{BDx0}
\end{equation}
where $0^{+}$ is a positive infinitesimal and $M_{ij}$ is the $2 \times 2$ matrix given by
\begin{equation}
M_{ij} = \dfrac{1}{t'}
\begin{pmatrix}
t t'-r r'& \sqrt{\dfrac{v_j}{v_i}} r' e^{i \varphi} \\
-\sqrt{\dfrac{v_i}{v_j}} r  e^{-i \varphi}&  1
\end{pmatrix}
\end{equation} 
with $\varphi=((k_{F1}+ k_{F2})+(\lambda_1+\lambda_2)\epsilon/L) x_0.$
The reflection and transmission coefficients are determined by  
\begin{align}
t e^{-i u_a}  &= t' e^{i u_a}  = \left(\cos d + i u_s\dfrac{\sin d}{d} \right)^{-1}, \nonumber\\
 r e^{-i \varphi} &=  r' e^{i \varphi}= -i  \sqrt{u_1 u_2}\frac{\sin d}{d} 
\cos \left(\frac{\theta_1-\theta_2}{2}\right) \sqrt{t t'}, \nonumber\\
d &= \frac{1}{2}\sqrt{u^2_1+u^2_2 -  2 u_1 u_2 \cos(\theta_1-\theta_2)},
\label{TandR}
\end{align}
where $v_0 = \hbar v_1 v_2/U_0$, $u_j=v_j/v_0,$ $u_s = (u_1+u_2)/2,$ and $u_a = (u_1-u_2)/2$. From the continuity conditions at $x=\pm L/2$ and Eq. \eqref{BDx0} we find the transcendental equation \eqref{single-barrier} where $\tau = |t|^2$. {As already noticed in the main text, Eq. \eqref{single-barrier} leads to split ABS when $v_1 \ne v_2$, except for $\delta=0,\pi$ where the ABS degeneracy is protected by time-reversal symmetry.}

\textbf{Double barrier model.} In this case, there are two boundary conditions similar as Eq.\,\eqref{BDx0} at the NS interfaces, which results in the transcendental equation 
\begin{align}
\sin(\tilde{\epsilon}_1 - \arccos \epsilon) \sin(\tilde{\epsilon}_2 - \arccos \epsilon)
=&(2-\tau_L-\tau_R) \sin(\tilde{\epsilon}_1) \sin(\tilde{\epsilon}_2) \nonumber\\
&-(1-\tau_L) (1-\tau_R) \sin(\tilde{\epsilon}_1 + \arccos \epsilon) \sin(\tilde{\epsilon}_2 + \arccos \epsilon) \nonumber\\
&-2 \sqrt{(1-\tau_L) (1-\tau_R)} \cos(\varphi_{\rm{tot}}) (1-\epsilon^2), 
\label{double-barrier}
\end{align} 
where $\tilde{\epsilon}_j=\epsilon \lambda_j+(-1)^j s \delta/2$, $\tau_{L,R}$ are the transmission probabilities at $x=\mp L/2,$ $\theta_{\nu}$ are the scattering phases acquired at the interfaces ($\nu\equiv L,R$):
\begin{equation}
\theta_{\nu} = \arg\left(\cos d_{\nu} + i \frac{\sin d_{\nu}}{d_{\nu}} \frac{v_s }{v_{\nu}} \right),
\end{equation}
where $d_\nu$ and $v_{\nu}$ are defined as $d$ in Eq. (\ref{TandR}) replacing $U_0$ by $U_\nu$. Finally, we note $\varphi_{\rm{tot}}=(k_{F1}+k_{F2}) L - (\theta_{L}+\theta_{R})$ the total accumulated phase. For the estimations quoted in the main text we have assumed two identical barriers, \textit{i.e.} $\tau_L=\tau_R=\tau$.

\begin{figure}[h]
\includegraphics[width=\columnwidth]{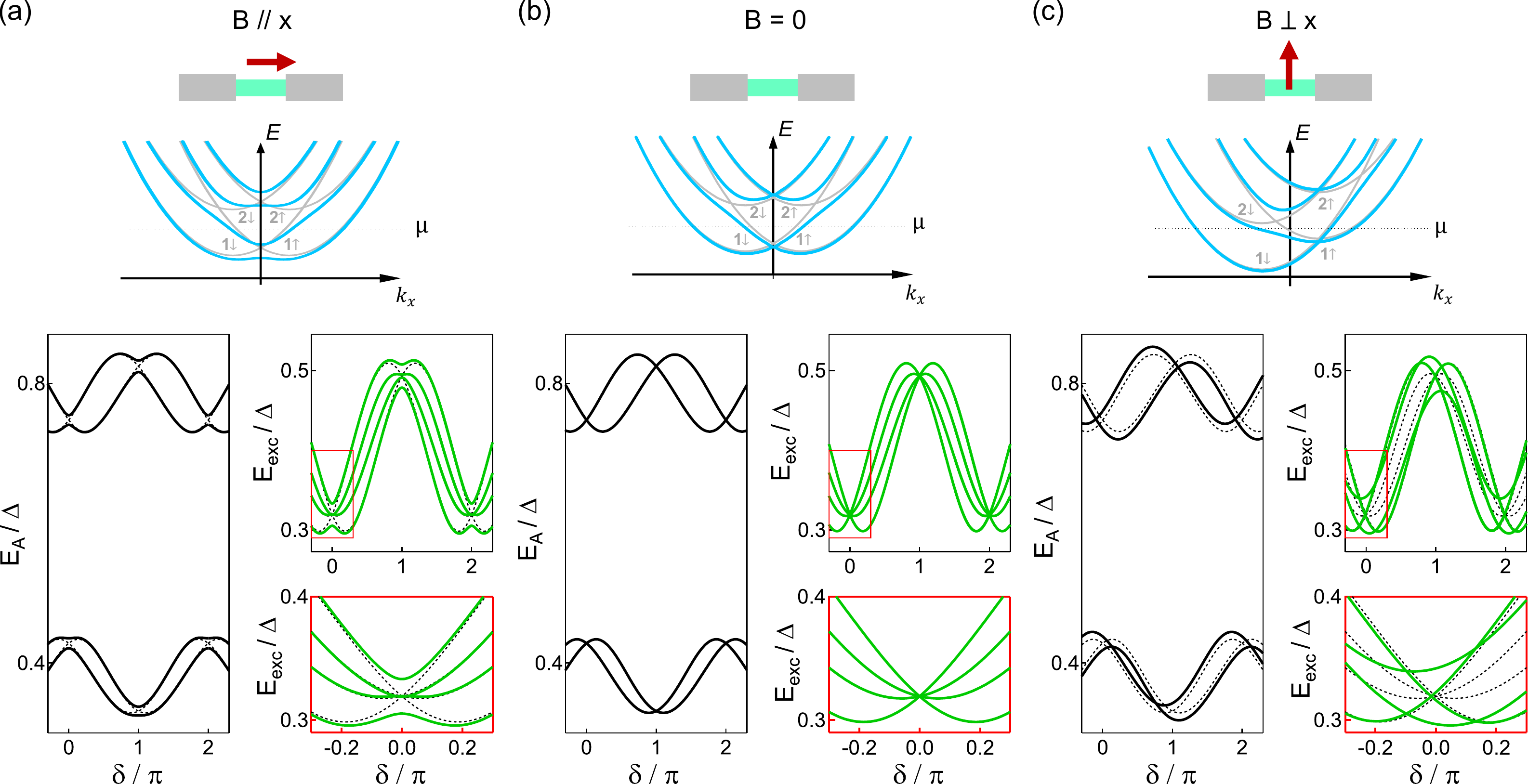}
\caption{Effect of an in-plane magnetic field on the band structure (top row), the Andreev levels (bottom row, left) and the excitation spectrum (bottom row, right). (b) reference graphs at zero field; (a) field applied along the wire axis; (c) field applied perpendicularly to the wire axis. 
{The field effect on the band structure is exaggerated for clarity. The model parameters for the Andreev levels and the excitation spectrum are the same as in Fig.\,\ref{Fig:Spectrum2} and $B = 10$ mT.}
}
\label{Fig:FieldBands}
\end{figure}

\textbf{Magnetic field effect.}
Information on the ABS spin structure can be inferred from their behavior in the presence of a finite magnetic field. This behavior depends strongly on the orientation of the field with respect to the nanowire axis \cite{Park2017}. 
We consider a magnetic field lying in the $xy$-plane. The $y$-component $B_y$ (parallel to the spin states of the transverse subbands without RSO) shifts the energy of the subbands depending on the spin states and modifies the Fermi wave vectors as illustrated in Fig.\,\ref{Fig:FieldBands}(c). They thus satisfy 
\begin{align}
E_s(k_{F}) = \frac{\hbar^2 k^2_F}{2 m^*} + \frac{E^{\perp}_1 +E^{\perp}_2}{2} 
- \sqrt{\left[  \frac{E^{\perp}_1 - E^{\perp}_2}{2} -s \left(\alpha k_F - \frac{g \mu_B}{2} B_y\right)\right]^2 + \eta^2} =\mu.
\end{align}

On the other hand, the $x$-component $B_x$ mixes opposite-spin states thus opening a gap at the crossings points as illustrated in Fig.\,\ref{Fig:FieldBands}(a). We include this effect perturbatively \cite{Park2017}. For both, $B\parallelsum x$ and $B\perp x$ cases, the resulting ABS and the corresponding transition lines are shown in the middle and bottom rows of Fig.\,\ref{Fig:FieldBands}.

\textbf{Fitting strategy}
The transcendental equations (Eqs.\,\eqref{single-barrier} and \eqref{double-barrier}) for the single and double barrier models contain dimensionless parameters with which we fit the experimental data {at zero magnetic field:}
\begin{itemize}
\item $\lambda_1$, $\lambda_2$, $\tau$, and $x_r$ for the single barrier model, 

\item $\lambda_1$, $\lambda_2$, $\tau$, and $\varphi_{\rm{tot}}$ for the double barrier model.
\end{itemize}    
We then deduce the physical parameters, $\alpha$, $\mu$ (measured from the bottom of the lowest band), $L$, and $U_0$ (or $U_{L/R}$) using Eqs.\,\eqref{Dispersion}, \eqref{Velocity} and \eqref{TandR}, and assuming that the nanowire diameter is fixed at $W=140$ nm.
We further set $m^*=0.023\,m_e$ where $m_e$ is the bare electron mass. For the experimental data in Fig.\,\ref{Fig:Spectrum2}, the single barrier model gives 
$\lambda_1=1.3,~\lambda_2=2.3,~\tau=0.295$, and $x_r=0.52$, resulting in the microscopic parameters  $\alpha=53\,\text{meV\,nm},~\mu=255~\mu\text{eV},~U_0=92\,\text{meV\,nm},~L=332\,\text{nm}.$ Using the double barrier model, we get $\lambda_1=1.1,~\lambda_2=1.9,~\tau=0.52,~\varphi_{\rm{tot}}=0.93~(\text{Mod}~2 \pi),~\alpha=36\,\text{meV\,nm},\,\mu=427\,\mu\text{eV},~U_L=U_R=130\,\text{meV\,nm},~L=314\,\text{nm}$. Another possibility is to fix the length of the junction $L$ to the length of the uncovered section of the InAs nanowire, $370\ $nm, which leads to $\alpha=38\ $meV nm and $\mu=422\,\mu$eV for the single barrier model ($\alpha=32\,$meV nm and $\mu=580\,\mu$eV for the double barrier model). However, in the single barrier model one cannot find values of $U_{0}$ leading to the corresponding $\tau$. This is due to the fact that in our simplified model for the scattering matrix, processes involving the higher subbands are neglected, thus limiting its validity to small values of $U_0$.

\begin{figure}[h]
\includegraphics[width=0.35\columnwidth]{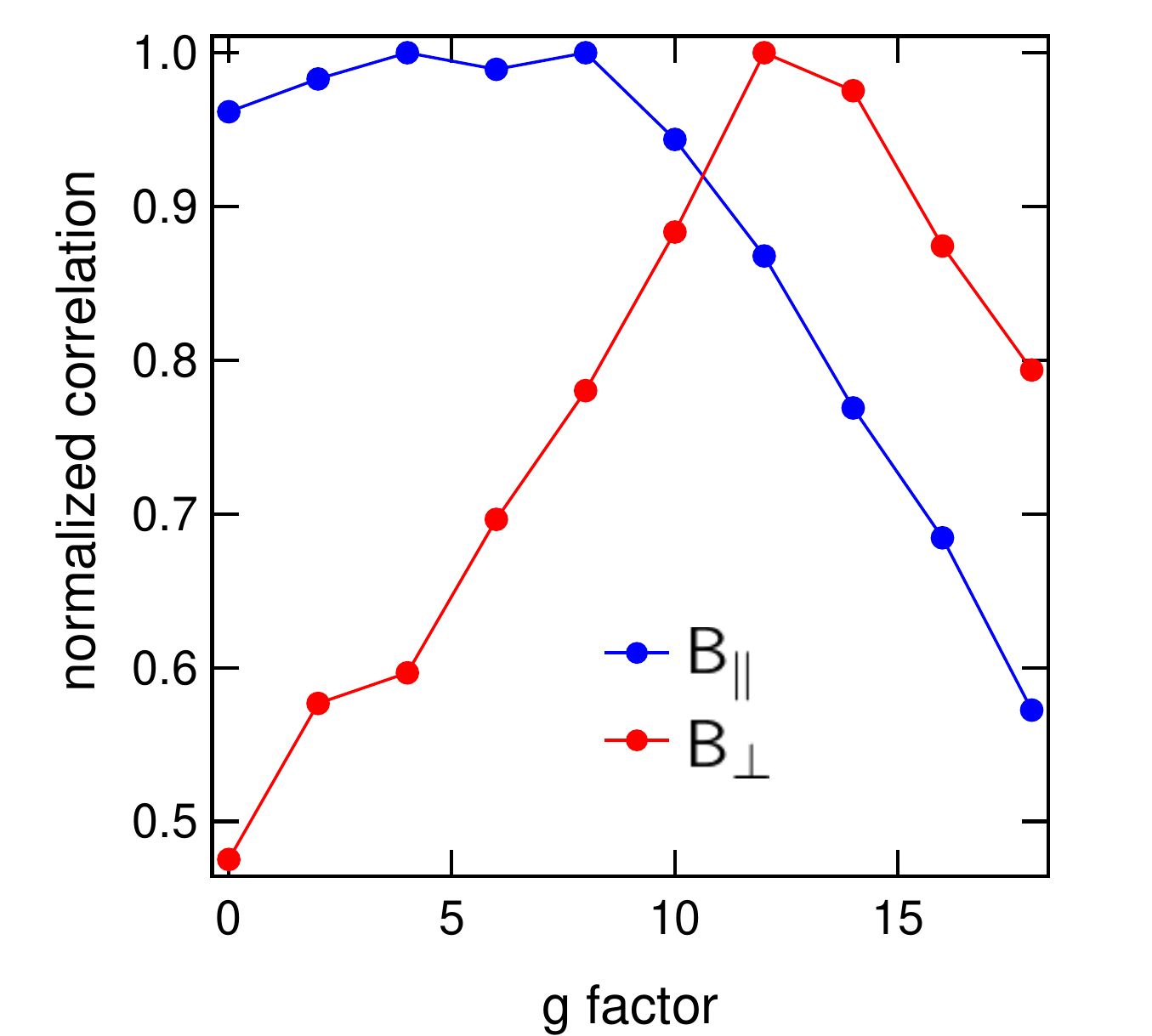}
\caption{{Correlation (normalized to maximum value) between finite magnetic field data and theory as a function of g factor, for field direction parallel (blue) or perpendicular (red) to the nanowire.}}
\label{correlation}
\end{figure}

{In order to fit the finite magnetic field data, in addition to the parameters determined at zero magnetic field, one needs the $g$-factors in the parallel and perpendicular directions, $g_{\parallelsum}$ and $g_{\perp}.$ We used all the data taken with field in the parallel and in the perpendicular directions, and calculated the correlation function between the images of the measured spectra (taking the absolute value of the response $f-f_0$) and theory, using various values of $g_{\parallelsum}$ and $g_{\perp}.$ Figure~\ref{correlation} shows the dependence of the correlation functions with $g_{\parallelsum}$ and $g_{\perp}.$ Best agreement is found for $g_{\parallelsum}=8$ and $g_{\perp}=12$, which are within the range of values reported in the literature \cite{Bjork2005,Deacon2011,dHollosy2013, Vaitiekenas2018}. Note that the determination of $g_{\parallelsum}$ is less accurate, and that, overall, $g_{\parallelsum}=4$ gives a similar correlation as $g_{\parallelsum}=8,$ but agreement is worse at largest values of $B_{\parallelsum}$ where the effect is the strongest.}

{\subsection{A2. Fit of the data at $V_g=-0.89~$V}}
{Many features of the data taken at $V_g=-0.89~$V (Fig.~\ref{Fig2:SpectrumZeroField}) can be accounted for by the single-barrier model. This is shown in Fig.~\ref{Fig8:SpectrumZeroField}, where we compare the data with the results of theory using $\lambda_1=2.81,$ $\lambda_2=4.7,$ $\tau=0.25$ and $x_r=0.17$. The Andreev spectrum obtained with this set of parameters (Fig.~\ref{Fig8:SpectrumZeroField}(c)) presents 3 manifolds of spin-split states, leading to 3 bundles of 4 lines associated to single-particle transitions between manifolds (green lines in Fig.~\ref{Fig8:SpectrumZeroField}(b)). They are in good agreement with transition lines at least partly visible in the data. In addition, the pair transition corresponding to two quasiparticles excited in the lowest manifold gives rise to an even transition which falls in the frequency range of the data, and roughly corresponds to a transition visible in the data. 
Assuming a fixed length $L=370$ nm and using the model of Eq.~\eqref{ModelHamiltonian}
we deduce the microscopic parameters $\alpha = 43.7$~meV~nm and $\mu = 102~\mu\text{eV}$ (measured from the bottom of the band). However, these values should be taken with care since the linearization of the dispersion relation is not valid for energies close to $\Delta$ when $\mu \lesssim \Delta$.}

\begin{figure}[h]
\includegraphics[width=0.8\columnwidth]{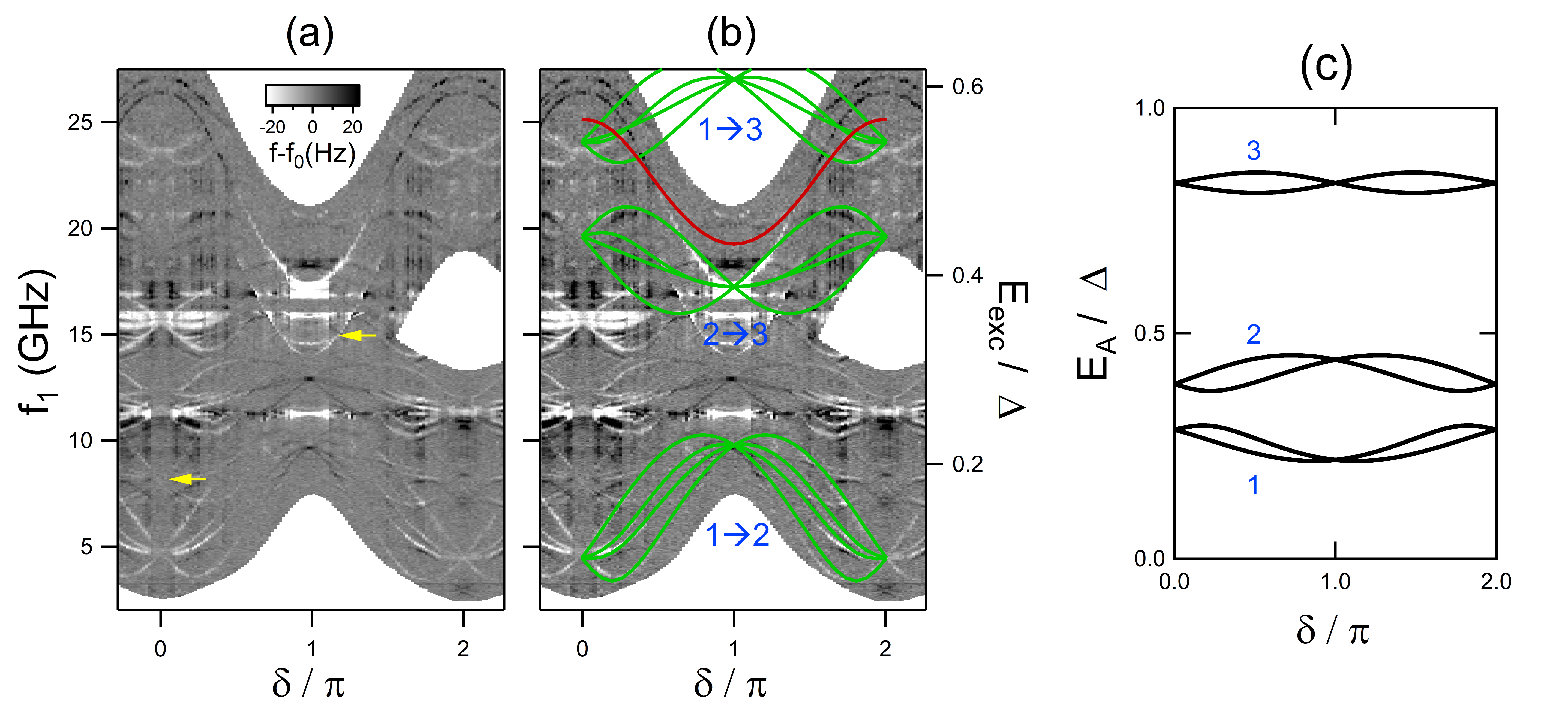}
\caption{{(a) Data at $V_g=-0.89~$V, with yellow arrows pointing to transition lines that are replicas of lines appearing exactly 3.26~GHz above. (b) Same data superimposed with predictions of the single barrier model, using parameters corresponding to the spectrum of ABS shown in (c). Single-particle transitions (green lines) between the three manifolds, labelled 1,2,3 in (c) are visible in the data. Red line in (b) is the pair transition leading to two quasiparticles in manifold 1.}}
\label{Fig8:SpectrumZeroField}
\end{figure}

{\subsection{A3. Measurement calibration}}
{The measurement is performed by chopping with a square-wave the excitation signal applied on the gate and recording with lock-in detectors the corresponding modulation of the response of the circuit on the two quadratures $I$ and $Q.$ We interpret these modulations as arising from shifts of the resonator frequency. To calibrate this effect, we measured how the DC values of $I$ and $Q$ change for small variations of the measurement frequency $f_0$ around 3.26~GHz. All the measurement chain being taken into account, we found $\frac{\partial I}{\partial f_0}=-40.3~\mu$V/Hz and $\frac{\partial Q}{\partial f_0}=34.4~\mu$V/Hz. The signal received by the lock-in measuring the $I$ quadrature is a square-wave, so that the response $I_{\rm{LI}}$ at the chopping frequency is related to the root mean square ($I_{\rm{rms}}$) and peak-to-peak ($I_{\rm{pp}}$) amplitudes at its input by $I_{\rm{LI}}=\frac{4}{\pi}I_{\rm{rms}}=\frac{\sqrt{2}}{\pi}I_{\rm{pp}}$. The same reasoning applies to the $Q$ quadrature measurement. We combine $I_{\rm{LI}}$ and $Q_{\rm{LI}}$ into $X_{\rm{LI}}=-\frac{I_{\rm{LI}}}{40.3}+\frac{Q_{\rm{LI}}}{34.4}$ and, using $\frac{\partial X}{\partial f_0}=2~\mu$V/Hz, the resonator frequency change $f-f_0$ is obtained from $f-f_0=\Delta f_0=\Delta X / (2~\mu \textrm{V/Hz})=\frac{X_{\rm{LI}}/(2~\mu \textrm{V/Hz})}{\sqrt{2}/\pi}$.}

{\subsection{A4. Gate dependence of the spectrum}}
\begin{figure*}[h!]
\includegraphics[width=0.8\columnwidth]{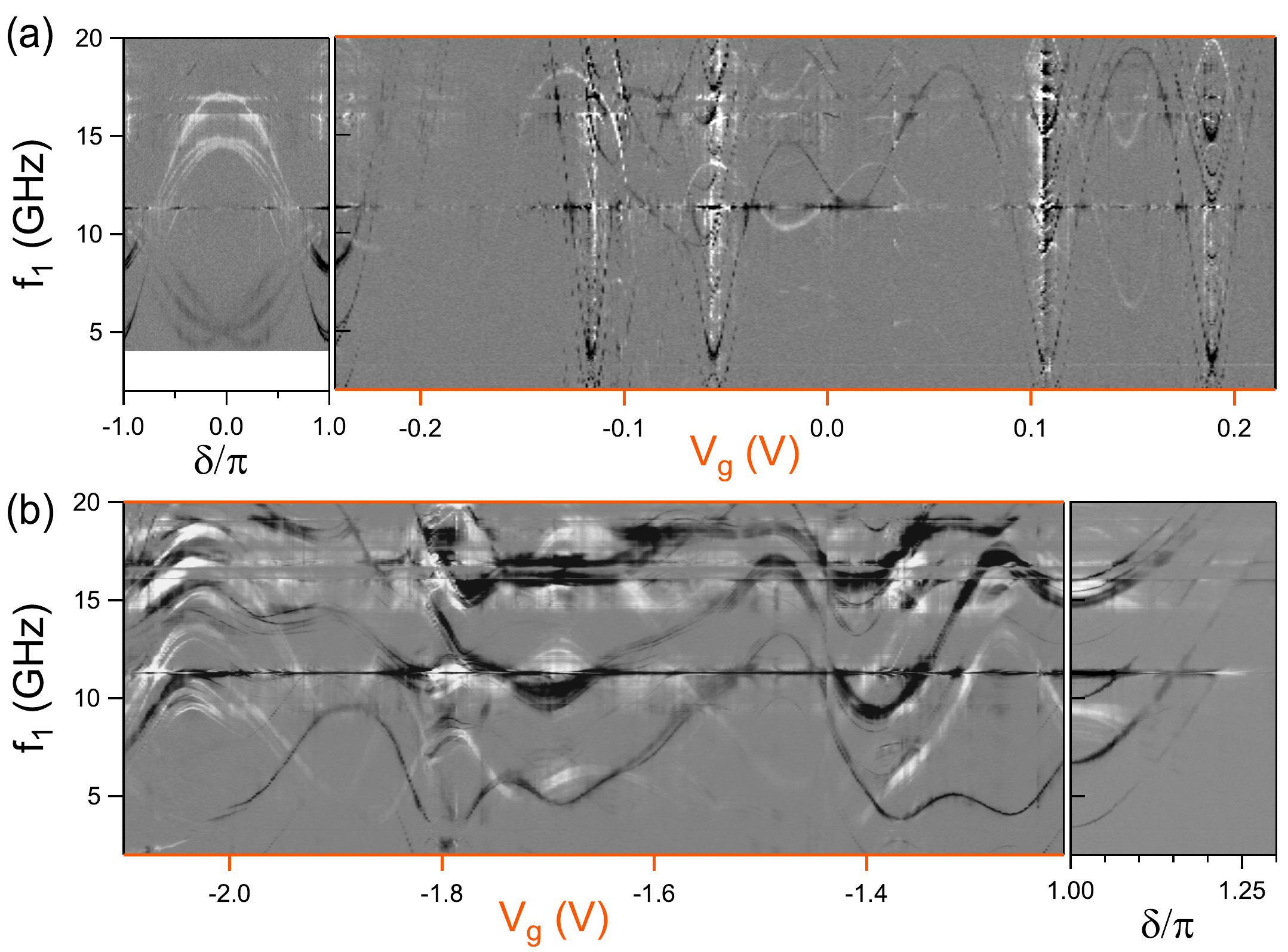}
\caption{Spectrum as a function of gate voltage at $\delta=\pi$ (right of (a), left of (b)), for two intervals of gate voltage, and taken during different cool-downs. (a) left panel and (b) right panel: phase dependence at the gate voltages corresponding to $V_g=-0.24~$V (a) (resp. $V_g=-1.21~$V (b)), which correspond to the leftmost (resp. rightmost) gate voltages of the panels showing the $V_g$ dependence.}
\label{Fig:SpectrumGateMap}
\end{figure*}

{Figure~\ref{Fig:SpectrumGateMap} shows two examples of the gate voltage-dependence of the spectrum at phase $\delta=\pi,$ with reference spectra as a function of phase. In both spectra, single-particle transitions appear white at $\delta=\pi,$ whereas pair transitions appear black. When $V_g$ is changed, both types of lines move up and down, but do not change color. Both types of transitions are observed in the frequency window 2-20~GHz at almost all values of $V_g$. A remarkable feature is that black and white lines move ``out of phase'', which can be understood from the effect of $V_g$ on the transmission $\tau$: when $\tau$ decreases, the distance between the two lowest manifolds decreases at $\delta=\pi,$ so that the transition energy for single-particle transitions decreases; at the same time, the energy of the lowest manifold increases, and so does the transition energy for pair transitions.
}

\end{widetext}


\begin{thebibliography}{50}

\bibitem{Golubev2004} A. A. Golubov, M. Y. Kupriyanov, and E. Ilichev, ``The current-phase relation in Josephson junctions'', Rev. Mod. Phys. \textbf{76}, 411 (2004).
\bibitem{Kulik1970} I.O. Kulik, Sov. Phys. JETP \textbf{30}, 944 (1970).
\bibitem{Beenakker1991a} C. W. J. Beenakker and H. van Houten, ``Josephson supercurrent through a superconducting quantum point contact shorter than the coherence length'',
Phys. Rev. Lett. \textbf{66}, 3056 (1991). 
\bibitem{Furusaki1991} A. Furusaki and M. Tsukada, ``Dc Josephson effect and Andreev reflection'', Solid State Commun. \textbf{78}, 299 (1991).
\bibitem{Bagwell1992} P. F. Bagwell,  ``Suppression of the Josephson current through a narrow, mesoscopic, semiconductor channel by a single impurity'', Phys. Rev. B \textbf{46}, 12573 (1992).
\bibitem{Pillet2010} J-D. Pillet, C. H. L. Quay, P. Morfin, C. Bena, A. Levy Yeyati, and P. Joyez, 
``Revealing the electronic structure of a carbon nanotube carrying a supercurrent'',
Nat. Phys. \textbf{6}, 965 (2010). 
\bibitem{Bretheau2013} L. Bretheau, \c{C}. \"{O}. Girit, H. Pothier, D. Esteve, and C. Urbina, ``Exciting Andreev pairs in a superconducting atomic contact'',
Nature \textbf{499}, 312 (2013).
\bibitem{Bretheau2013_2} L. Bretheau, \c{C}. \"{O}. Girit, C. Urbina, D. Esteve, and H. Pothier, ``Supercurrent spectroscopy of Andreev states'',
Phys. Rev. X \textbf{3}, 041034 (2013).
\bibitem{Janvier2015} C. Janvier, L. Tosi, L. Bretheau, \c{C}. \"{O}. Girit, M. Stern, P. Bertet, P. Joyez, D. Vion, D. Esteve, M. F. Goffman, H. Pothier, and C. Urbina, 
``Coherent manipulation of Andreev states in superconducting atomic contacts'', Science \textbf{349}, 1199 (2015).
\bibitem{Lee2014} E. J. H. Lee, X. Jiang, M. Houzet, R. Aguado, C. M. Lieber, and S. De Franceschi, ``Spin-resolved Andreev levels and parity crossings in hybrid superconductor-semiconductor nanostructures'', Nat. Nanotech. \textbf{9}, 79 (2014).
\bibitem{Woerkom2016} D. J. van Woerkom, A. Proutski, B. van Heck, Dani\"{e}l Bouman, J. I. V\"{a}yrynen, L. I. Glazman, P. Krogstrup, J. Nyg\r{a}rd, L. P. Kouwenhoven, Attila Geresdi, ``Microwave spectroscopy of spinful Andreev bound states in ballistic semiconductor Josephson junctions'', Nature Physics \textbf{13}, 876 (2017).
\bibitem{Hays2017} M. Hays, G. de Lange, K. Serniak, D. J. van Woerkom, D. Bouman, P. Krogstrup, J. Nyg\r{a}rd, A. Geresdi, and M. H. Devoret, ``Direct microwave measurement of Andreev-bound-state dynamics in a proximitized semiconducting nanowire'', Phys. Rev. Lett. \textbf{121}, 047001 (2018).
\bibitem{Michelsen2008} J. Michelsen, V. S. Shumeiko, and G. Wendin, ``Manipulation with Andreev states in spin active mesoscopic Josephson junctions'', Phys. Rev. B \textbf{77}, 184506 (2008).
\bibitem{DeFranceschi2010} S. De Franceschi, L. P. Kouwenhoven, C. Sch\"{o}nenberger, and W. Wernsdorfer, ``Hybrid superconductor - quantum dot devices'', Nat. Nanotech. \textbf{5}, 703 (2010).
\bibitem{Linder2015} Jacob Linder and Jason W. A. Robinson, ``Superconducting spintronics'', Nature Physics \textbf{11}, 307 (2015).
\bibitem{Prada2017} Elsa Prada, Ram\'{o}n Aguado, and Pablo San-Jose, ``Measuring Majorana nonlocality and spin structure with a quantum dot'', Phys. Rev. B \textbf{96}, 085418 (2017).
\bibitem{Zazunov2017} A. Zazunov, R. Egger, M. Alvarado, and A. Levy Yeyati, ``Josephson effect in multiterminal topological junctions'', Phys. Rev. B \textbf{96}, 024516 (2017).
\bibitem{Deng2018} M. T. Deng, S. Vaitiekenas, E. Prada, P. San-Jose, J. Nyg\r{a}rd, P. Krogstrup, R. Aguado, and C. M. Marcus, ``Nonlocality of Majorana modes in hybrid nanowires'', Phys. Rev. B \textbf{98}, 085125 (2018).
\bibitem{Hart2017} Sean Hart, Hechen Ren, Michael Kosowsky, Gilad Ben-Shach, Philipp Leubner, Christoph Br\"une, Hartmut Buhmann, Laurens W. Molenkamp, Bertrand I. Halperin and Amir Yacoby, ``Controlled finite momentum pairing and spatially varying order parameter in proximitized HgTe quantum wells'', Nat. Phys. \textbf{13}, 87 (2017).
\bibitem{Chtchelkatchev2003} N. M. Chtchelkatchev and Y. V. Nazarov, ``Andreev quantum dots for spin manipulation'',
Phys. Rev. Lett. \textbf{90}, 226806 (2003). 
\bibitem{Padurariu2010} C. Padurariu and Y. V. Nazarov, ``Theoretical proposal for superconducting spin qubits'',
Phys. Rev. B \textbf{81}, 144519 (2010).
\bibitem{Beri2008} B. B\'{e}ri, J. H. Bardarson, and C. W. J. Beenakker, ``Splitting of Andreev levels in a Josephson junction by spin-orbit cbezuglioupling'', Phys. Rev. B \textbf{77}, 045311 (2008).
\bibitem{Cayao2015} Jorge Cayao, Elsa Prada, Pablo San-Jose, and Ram\'{o}n Aguado, ``SNS junctions in nanowires with spin-orbit coupling: Role of confinement and helicity on the subgap spectrum'', Phys. Rev. B \textbf{91}, 024514 (2015).
\bibitem{Heck2017} B. van Heck, J. I. V\"{a}yrynen, and L. I. Glazman, ``Zeeman and spin-orbit effects in the Andreev spectra of nanowire junctions'',
Phys. Rev. B \textbf{96}, 075404 (2017).
\bibitem{Bychkov1984} Yu. A. Bychkov and E. I. Rashba, ``Properties of a 2D electron gas with lifted spectral degeneracy'', JETP Letters \textbf{39} 78 (1984).
\bibitem{Park2017} Sunghun Park and A. Levy Yeyati, ``Andreev spin qubits in multichannel Rashba nanowires'', Phys. Rev. B \textbf{96}, 125416 (2017).
\bibitem{Reynoso2012} A. A. Reynoso, G. Usaj, C. A. Balseiro, D. Feinberg, and M. Avignon,
``Spin-orbit-induced chirality of Andreev states in Josephson junctions'',
Phys. Rev. B \textbf{86}, 214519 (2012). 
\bibitem{Yokoyama2014} T. Yokoyama, M. Eto, and Y. V. Nazarov, 
``Anomalous Josephson effect induced by spin-orbit interaction and Zeeman effect in semiconductor nanowires'',
Phys. Rev. B \textbf{89}, 195407 (2014).
\bibitem{Murani2016} A. Murani, A. Chepelianskii, S. Gu\'{e}ron, and H. Bouchiat, ``Andreev spectrum with high spin-orbit interactions: revealing spin splitting and topologically protected crossings'', Phys. Rev. B \textbf{96}, 165415 (2017). 
\bibitem{Moroz1999} A. V. Moroz and C. H. W. Barnes, ``Effect of the spin-orbit interaction on the band structure and conductance of quasi-one-dimensional systems'', Phys. Rev. B \textbf{60}, 14272 (1999).
\bibitem{Governale2002} M. Governale and U. Z\"ulicke, ``Spin accumulation in quantum wires with strong Rashba spin-orbit coupling'', Phys. Rev. B \textbf{66}, 073311 (2002).
\bibitem{Zgirski2011} M. Zgirski, L. Bretheau, Q. Le Masne, H. Pothier, D. Esteve, and C. Urbina, 
``Evidence for long-lived quasiparticles trapped in superconducting point contacts'',
Phys. Rev. Lett. \textbf{106}, 257003 (2011).
\bibitem{Vayrynen2015} Jukka I. V\"{a}yrynen, Gianluca Rastelli, Wolfgang Belzig, and Leonid I. Glazman, Phys. Rev. B \textbf{92}, 134508 (2015).
\bibitem{Krogstrup2015} P. Krogstrup, N. L. B. Ziino, W. Chang, S. M. Albrecht, M. H. Madsen, E. Johnson, J. Nyg\r{a}rd, C. M. Marcus and T. S Jespersen., ``Epitaxy of semiconductor-superconductor nanowires'', Nat. Mater. \textbf{14}, 400 (2015).
\bibitem{Chang2015} W. Chang, S. M. Albrecht, T. S. Jespersen, F. Kuemmeth, P. Krogstrup, J. Nyg\r{a}rd and C. M. Marcus, ``Hard gap in epitaxial semiconductor-superconductor nanowires'', Nature Nanotechnology \textbf{10}, 232 (2015).
\bibitem{Goffman2017} M. F. Goffman, C. Urbina, H. Pothier, J. Nyg\r{a}rd, C. M. Marcus, and P. Krogstrup, 
``Conduction channels of an InAs-Al nanowire Josephson weak link'', New J. Phys. \textbf{19}, 092002 (2017).
\bibitem{Samuelsson2000} P. Samuelsson, J. Lantz, V. S. Shumeiko, and G. Wendin, ``Nonequilibrium Josephson effect in mesoscopic ballistic multiterminal SNS junctions'', Phys. Rev. B \textbf{62}, 1319 (2000).
\bibitem{Fasth2007} C. Fasth, A. Fuhrer, L. Samuelson, Vitaly N. Golovach, and Daniel Loss, ``Direct Measurement of the Spin-Orbit Interaction in a Two-Electron InAs Nanowire Quantum Dot'', Phys. Rev. Lett. \textbf{98}, 266801 (2007).
\bibitem{Scherubl2016} Zolt\'an Scher\"ubl, Gergo F\"ul\"op, Morten H. Madsen, Jesper Nyg\r{a}rd, and Szabolcs Csonka, ``Electrical tuning of Rashba spin-orbit interaction in multigated InAs nanowires'', Phys. Rev. B \textbf{94}, 035444 (2016).
\bibitem{angle} The perfect symmetry of the spectrum when the field is applied parallel to the wire was used to define precisely the field angle, in agreement within a few degrees with a determination from the images of the sample.
\bibitem{Degtyarev2017} V. E. Degtyarev, S. V. Khazanova and N. V. Demarina, ``Features of electron gas in InAs nanowires imposed by interplay between nanowire geometry, doping and surface states'', Scientific Reports \textbf{7}, 3411 (2017).
\bibitem{Zuo2017} Kun Zuo, Vincent Mourik, Daniel B. Szombati, Bas Nijholt, David J. van Woerkom, Attila Geresdi, Jun Chen, Viacheslav P. Ostroukh, Anton R. Akhmerov, Sebastién R. Plissard, Diana Car, Erik P. A. M. Bakkers, Dmitry I. Pikulin, Leo P. Kouwenhoven, and Sergey M. Frolov, ``Supercurrent Interference in Few-Mode Nanowire Josephson Junctions'', Phys. Rev. Lett. \textbf{119}, 187704 (2017).
\bibitem{Antipov2018} Andrey E. Antipov, Arno Bargerbos, Georg W. Winkler, Bela Bauer, Enrico Rossi, and Roman M. Lutchyn,``Effects of Gate-Induced Electric Fields on Semiconductor Majorana Nanowires'', Phys. Rev. X \textbf{8}, 031041 (2018)
\bibitem{Winkler2018} Georg W. Winkler, Andrey E. Antipov, Bernard van Heck, Alexey A. Soluyanov, Leonid I. Glazman, Michael Wimmer, Roman M. Lutchyn, ``A unified numerical approach to semiconductor-superconductor heterostructures'', arXiv:1810.04180 (2018).
\bibitem{Gorelik1995} L. Y. Gorelik, V. S. Shumeiko, R. I. Shekhter, G. Wendin, and M. Jonson, ``Microwave-Induced "Somersault Effect" in Flow of Josephson Current through a Quantum Constriction'', Phys. Rev. Lett. \textbf{75}, 1162 (1995)
\bibitem{Desposito2001} M. A. Desp\'{o}sito and A. Levy Yeyati, ``Controlled dephasing of Andreev states in superconducting quantum point contacts'', Phys. Rev. B \textbf{64}, 140511(R) (2001).
\bibitem{Quay2015} C.H.L. Quay, M. Weideneder, Y. Chiffaudel, C. Strunk, and M. Aprili, Nat. Commun. \textbf{6}, 8660 (2015).
\bibitem{Bjork2005} M. T. Bj\"ork, A. Fuhrer, A. E. Hansen, M. W. Larsson, L. E. Fr\"{o}berg, L. Samuelson, ``Tunable effective g-factor in InAs nanowire quantum dots'', Phys. Rev. B \textbf{72}, 201307 (2005). 
\bibitem{Deacon2011} R. S. Deacon, Y. Kanai, S. Takahashi, A. Oiwa, K. Yoshida, K. Shibata, K. Hirakawa, Y. Tokura, and S. Tarucha,``Electrically tuned g tensor in an InAs self-assembled quantum dot'', Phys. Rev. B \textbf{84}, 041302(R) (2011).
\bibitem{dHollosy2013} Samuel d'Hollosy, Gábor Fábián, Andreas Baumgartner, Jesper Nygård, Christian Schönenberger, ``g-factor anisotropy in nanowire-based InAs quantum dots'', AIP Conf. Proc. \textbf{1566}, 359 (2013).
\bibitem{Vaitiekenas2018} S. Vaitiekenas, M.-T. Deng, J. Nyg\r{a}rd, P. Krogstrup, and C. M. Marcus, ``Effective g Factor of Subgap States in Hybrid Nanowires'', Phys. Rev. Lett. \textbf{121}, 037703 (2018).
\end{thebibliography}
\end{document}